\documentclass[paper]{ieice}
\usepackage[pdftex]{graphicx,xcolor}
\usepackage{url}
\usepackage[fleqn]{amsmath}
\usepackage{newtxtext}
\usepackage[varg]{newtxmath}
\usepackage{multirow}

\field{B}
\title{Internet Anomaly Detection based on Complex Network Path}
\authorlist{%
\authorentry[jinfa.wong@gmail.com]{Jinfa WANG}{s}{A1}\MembershipNumber{1784226}
\authorentry{Siyuan Jia}{n}{A1}
\authorentry{Hai Zhao}{n}{A1}
\authorentry{Jiuqiang Xu}{n}{A1}
\authorentry{Chuan Lin}{n}{A1}
}
\affiliate[A1]{The authors are with the School of Computer Science and Engineering, Northeastern University, Shenyang, 110169 China.}
\received{2017}{11}{1}
\revised{2015}{1}{1}

\newtheorem{definition}{Definition}

\def\UrlOrds{\do\*\do\-\do\~\do\'\do\"\do\_\do\.}%
\usepackage{url}
\makeatletter
\g@addto@macro{\UrlBreaks}{\UrlOrds}
\makeatother

\begin{document}
\maketitle
\begin{summary}
Detecting the anomaly behaviors such as network failure or Internet intentional attack in the large-scale Internet is a vital but challenging task. While numerous techniques have been developed based on Internet traffic in past years, anomaly detection for structured datasets by complex network have just been of focus recently. In this paper, a anomaly detection method for large-scale Internet topology is proposed by considering the changes of network crashes. In order to quantify the dynamic changes of Internet topology, the network path changes coefficient(NPCC) is put forward which will highlight the Internet abnormal state after it is attacked continuously. Furthermore we proposed the decision function which is inspired by Fibonacci Sequence to determine whether the Internet is abnormal or not. That is the current Internet is abnormal if its NPCC is beyond the normal domain which structured by the previous $k$ NPCCs of Internet topology. Finally the new Internet anomaly detection method was tested over the topology data of three Internet anomaly events. The results show that the detection accuracy of all events are over 97\%, the detection precision of each event are 90.24\%, 83.33\% and 66.67\%, when $k = 36$. According to the experimental values of the index $F_1$, we found the the better the detection performance is, the bigger the $k$ is, and our method has better performance for the anomaly behaviors caused by network failure than that caused by intentional attack. Compared with traditional anomaly detection, our work may be more simple and powerful for the government or organization in items of detecting large-scale abnormal events.
\end{summary}
\begin{keywords}
Internet, Anomaly detection, Complex network, Network Diameter, Network effective path, Network mean shortest path
\end{keywords}

\section{Introduction}
\label{sec:introduction}
Internet is playing a key role in society, economy and human daily life nowadays, so that cyberspace security becomes one of the hottest topics. Indeed it is facing a series security problems, such as the malware and ransomware on personal devices, information leak and theft, intentional attack and irresistible natural disaster. Especially intentional attack for the targeted company, organization, government and even local Internet based on worm, DDoS, Spoofing and system vulnerabilities are becoming more and more common. With the rapid development of IoT(The Internet of Things), the "destruction of service"(DeOS) attack in the future will even interrupt the Internet itself which reported in the 2017 Mid-year Cybersecurity Report from Cisco \cite{cisco2017}. For such a large-scale network failure, how to build a more powerful method to detect them in global Internet in time is an vital task.

Anomaly detection is a method to find patterns that deviate from the expected behavior. These nonconforming patterns, occurring rarely in datasets, are often referred to as anomalies, outliers, exceptions, aberrations, surprises, or contaminates in the field of specific research. In a word, our studies are to warn the problems what are happening, or predict the system evolving trend for a long time by detecting the local or global unusual changes\cite{chandola2009,Bhuyan2014,Akoglu2015}.

The past works mainly focus on the Internet traffic anomaly, that is keeping traffic statistic such as packet size, and inter-arrivals, flow accounts, byte volumes, etc., analyzing the traffic behaviors, constructing the outliers model of the traffic, and finally giving the traffic clustering results: normal and abnormal\cite{Bhuyan2014,Uchida2012,Fontugne2011,Song2009,Matsuda2017}. However Iliofotou et.al\cite{Iliofotou2007} proposed the Traffic Dispersion Graph(TDG), which extracts on network-wide interactions of hosts, to monitor, analyze, and visualize network traffic. It is the first study which considers the inter-dependency of traffic data by structuring the network traffic data to complex network. Subsequently Le et.al\cite{Le2011} used  complex network concepts such as degree distribution, maximum degree and dK-2 distance to detect anomalous network traffic. In fact, detecting the anomalies in datasets depicted by dynamic network has received much attention in recent years, since dynamic network can provide a powerful machinery for effectively capturing these long-range correlations among inter-dependent data objects \cite{Akoglu2015}. For instance the intentional attack for Internet\cite{Trajanovski2012}. The methods from Refs.\cite{Bhuyan2014,Uchida2012,Fontugne2011,Song2009,Matsuda2017} would find the abnormal behaviors of the monitored network, if they have permission to analyze the incoming and outgoing network traffic. In other words there is no Internet anomaly detection capability without the permission of network traffic monitoring. But the Internet attack usually aims to the LAN such as the computer network of the company, organization, or government. One failing network device may result in its neighbors failure, i.e. not access the Internet, moreover the failing neighbors would influence its neighbors' neighbor. As a result, the attacked LAN will disappears in the routing tables of global Internet. So if we study this problem from the perspective of Internet topology structure, i.e, the dependencies between network devices(here one topology node represents a network device and the edges depict the dependencies), we not only eliminate the interference of noise nodes in non-attack area\cite{Krasichkov2015}, but also locate the abnormal areas in global Internet and even predict the trend of this event. Based on the complex network theory, researchers have intensified their study of methods for anomaly detection in structured network data, such as cyber networks\cite{Sequeira2002}, fraud detection\cite{Chau2006}, fault detection in medical claims, engineering systems\cite{Fujimaki2005}, sensor networks, climate network\cite{Guez2014} and many more domains. Therefore the works about Internet anomaly detection are very significant for researchers or engineers.

In this paper, we attempt to detect the Internet anomaly behaviors from the perspective of the dynamic of IP-level Internet topology structure. On one hand, the Internet topology describe the inter-dependencies between hundreds and thousands network devices such as router, server, computer, mobile and others. Then one topology snapshot of Internet must reflect its connections at that time. On the other hand, for the intentional attack or future destruction of service which will bring about the local network collapsing, building a system of monitoring global Internet has become an imminent task for researchers. In order to implement the purpose of Internet anomaly detection, there are three questions answered in the following sections: (1) why the Internet topology can accurately feedback Internet anomaly behaviors caused by intentional attack or network failure; (2) how to quantify the sudden changes of Internet topology structure; (3) how to effectively distinguish the abnormal and normal behaviors of Internet. Finally the Internet topology data of anomaly events, including 2011's Japanese earthquake, 2014's BGP table hitting 512k and 2016's American DDoS attack, are used to verify the effectiveness and performance of our method. This paper's contributions are as follows:
\begin{itemize}
\item
    Exporting the correlation between the dynamic of Internet topology and the anomaly behaviors, eg. network devices failure or Internet intentional attack. This is because the Internet topology consisted of traceroute paths is changing dynamically along with the global Internet environment.
\item
    Proposing the network path changing coefficient (NPCC) which describes the relative changing ratio among the network diameter, network effective path and network mean shortest path length, and reveals the abnormal changes ratio, i.e. first increase and then decrease, under the continuous network failure or attack.
\item
    A decision function to determine whether the Internet is abnormal or not. It forms a normal area by equation \eqref{eq:f-func} which is inspired by Fibonacci Sequence, i.e. the current network state will be related to previous $k$ sampled networks.
\item
    A effective and powerful methodology to detect the global Internet anomaly behaviors by monitoring the Internet topology in real time.
\end{itemize}

The remainder of this article is organized as follows. Section \ref{sec:related-work} surveys the works related to Internet anomaly detection and the evolution dynamics of Internet topology. In Section \ref{sec:internet-anomaly-phenomena}, we discuss the Internet topology structure anomaly due to the intentional attack and network failure. Then the Internet anomaly detection model based on network diameter is proposed in Section \ref{sec:internet-anomaly-detection-method}. We show our model performance and have a detailed discussion about the experimental results in Section \ref{sec:experimental-results}. Finally Section \ref{sec:conclusions} concludes this paper and gives our future work.

\section{Related Work}
\label{sec:related-work}

The anomaly-based intrusion detection refers to the problem of finding exceptional patterns in network traffic that do not conform to our excepted normal behavior. In the beginning of the study, Anderson\cite{Anderson1980} believed that an intrusion attempt or a threat is a deliberate and unauthorized attempt to access information, manipulate information, or render a system unreliable or unusable, and the network traffic would carry all traces in process of information access. Since then the network traffic becomes the best research object of network anomaly detection. For instance, Park et.al.\cite{Park2007} developed the unknown worms detection method by checking the random distribution of destination addresses in network traffic. In Ref.\cite{Ishibashi2008}, the authors found the heavy hitters in terms of cardinality in massive traffic, and introduced a new algorithm to anomaly detection. Furthermore, the researchers has studied the Internet anomaly detection based on pattern recognition\cite{Uchida2012,Fontugne2011}, SVM\cite{Song2009} and PCA\cite{Matsuda2017}.Compare with the above traffic statistic studies, Iliofotou et.al\cite{Iliofotou2007} proposed the Traffic Dispersion Graph(TDG) to discover network-wide interactions of hosts, to monitor, analyze, and visualize network traffic. It is the first study which considers the inter-dependency relationship of network traffic data. Subsequently Le et.al\cite{Le2011} used  complex network theory such as degree distribution, maximum degree and dK-2 distance to detect anomalous network traffic. However the study of network-wide anomaly detection based on routers' connecting relationships is presented by Zhou\cite{ZHOU2011}. He used the graph to describe the traffic feature distribution sequences and their relationships. Therefore the complex network or graph is an effective means to detect network anomaly.

Since the continuous increase in the size of Internet and more and more Internet services being transferred to cloud platform, one method, which can detect global Internet anomalies effectively, and be deployed easily, have to be introduced to accepting the new challenges. So in this paper we will introduce the real-time Internet topology structure to detect the abnormal behaviors which causes the sudden changes of Internet structure. As far as we know, no attention has been paid to detect network anomalies from the perspective of IP-level Internet topology. But the related research such as the dynamics of Internet topology, IP-level Internet probing, the phase transaction of Internet structure have begun.

Bourgeau\cite{Bourgeau2011} analyzed the network topology dynamism captured in a real measurement scenario and quantified the impact of coarser time granularity on the dynamism information missed. It helps us to understand network topology dynamical behavior by traceroute detection method. Planck et.al.\cite{Planck2011} presented a new framework for realizing near real-time global scale disruptive Internet event detection based on Hidden Markov Models. This framework consists of three steps: BGP data processing, automated event characterization, and result visualization. Ai et.al.\cite{Ai2013} analyzed IPv6 Internet topology evolution in IP-level topology to demonstrate how it changes in uncommon ways to restructure the Internet. Latapy et.al.\cite{Latapy2014} explored an empirical approach based on a notion of statistically significant events. It identifies some properties of graph dynamic, numbers of nodes, connected components, and distance, which exhibit a homogeneous distribution with outliers, corresponding to events. Then one method which compares current distribution fit to original distribution of graph properties is proposed to detect Internet events.

\section{Internet Anomaly Phenomena}
\label{sec:internet-anomaly-phenomena}

\subsection{Network failure and intentional attack}
\label{sec:internet-infrastructure-failure}

Internet Anomaly usually results from two aspects of reasons: one is the network device failure, for example the nature disease and war could make Internet lose the power, damage the routers, servers and cables so that interrupt the connectivities between the communication devices, in addition to the device aging; The other is the human's intentional attacks, which bring about the local network overload, and further influence the network service and network performance seriously.

Whatever the cause of the anomaly is, eventually two kinds of results may occur: communication connections breaking or routing table changing. Especially when one device fails, all connections to its neighbors are interrupted. Subsequently its pre-hop routers need to look for new router interfaces as next-hop which can route to destination address\cite{Singh2003}. Undoubtedly these new routing paths will be broadcast in its LAN, and then  the current Internet topology structure is changed. Thus the network failure not only remove some old connections, but also could reconstruct some new connections under the rerouting mechanism.

In general, when Internet is attacked intentionally, it accompanies by the surge of network traffic yet. At this time, if network load is close to network capacity, network throughput will increase slowly, but network delay is rising rapidly. As a result, the bigger network delay improves the probability of packet loss. If the increasing network load continuously is greater than network capacity, the network throughput will have a jumpingly decline. In this case, this device begins to adopt the congestion control\cite{Mamun-Or-Rashid2007} and router load balancing\cite{Singh2003} strategy to decrease the load of current device. Then the backup routes are enabled, or the new routing paths are constructed. However, for the Internet intentional attack behaviors, especially the DDoS, there are a large number of the request packets to be directed to the same target address suddenly. In this situation, the target device has to shutdown because its network load is far greater than its network capacity.

\subsection{Internet Topology Structure Anomaly}
\label{sec:internet-topology-anomaly}

Internet topology structure releases the routing relationships between network devices. Nowadays the Internet probe skills include active probe and passive probe. The active probe, such as Ark, Dimes, DipZoom, PlanetLab, Rocketfuel and iPlane, requires a router to send probe packet to other routers and expects to receive responses from them. Upon receiving a probe request message, a router is expected to create response packet and send it back to the probe originator. The passive probe, such as Route Views, RIPE NCC, IRR and IRL, tracks the performance and behavior of packet streams by monitoring the traffic without creating or modifying it, or collects the router table information from the authorized devices.

\begin{figure}[b]
\begin{center}
\includegraphics[width=0.8\columnwidth]{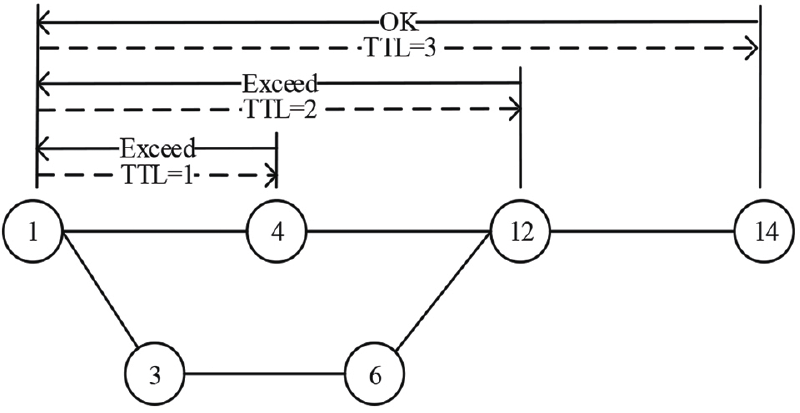}
\end{center}
\caption{Diagram of traceroute detection.}
\label{fig:traceroute}
\end{figure}

In this paper, we mainly focus on the dynamical changes of large-scale Internet topology structure and attempt to construct a method to quantify the abnormal and normal behaviors. So the active probe is better for our work. Then the below, as Fig.\ref{fig:traceroute} shown, describes the active probe process. There are two paths from router $R1$ to $R14$. We hypothesis that the routers adopt the shortest path selection routing protocol. Then one probe packet sent by router $R1$ will reach router $R14$ via routers $R4$ and $R12$. In this process, the router $R1$ sends an ICMP request packet with $TTL=1$(Time to Live) firstly, which its destination address is $R4$. Then the router $R4$ receives the packet from $R1$ and decrements the TTL value. If the current TTL value is 0, this packet dies and the router $R4$ is supposed to send back a ''Time Exceeded'' ICMP message. As this ICMP message contains the IP address of the router $R4$, the $R1$ records the $R4$'s address with $TTL=1$. Second, $R1$ creates an ICMP request packet with $TTL=2$ and gets next-hop address by analyzing the ''Time Exceeded'' ICMP message from router $R12$. Next $R1$ sends the request message with the incremental TTL value and extracts the next-hop IP address. When router $R1$ receives the ICMP ECHO message, this probe exits.

As we all know, after network devices fail or the Internet is attacked maliciously, not only the in- and out-connections of those devices are interrupted, but also the routing relationships are changed. For instance, the router $R1$ selects the router $R3$ as its next-hop interface to forward packets for the routers $R12$ or $R14$, when the router $R4$ is in the state of network overload and even fails. At this time executing traceroute again. Although we has obtained the routing path from $R1$ to $R14$, get the different Internet topology structure due to the fault of router $R4$. Therefore enough traceroute paths form the complex Internet topology, of course the network topology have to change along with device state and routing table.

\begin{figure}[b]
\centering
\includegraphics[width=\columnwidth]{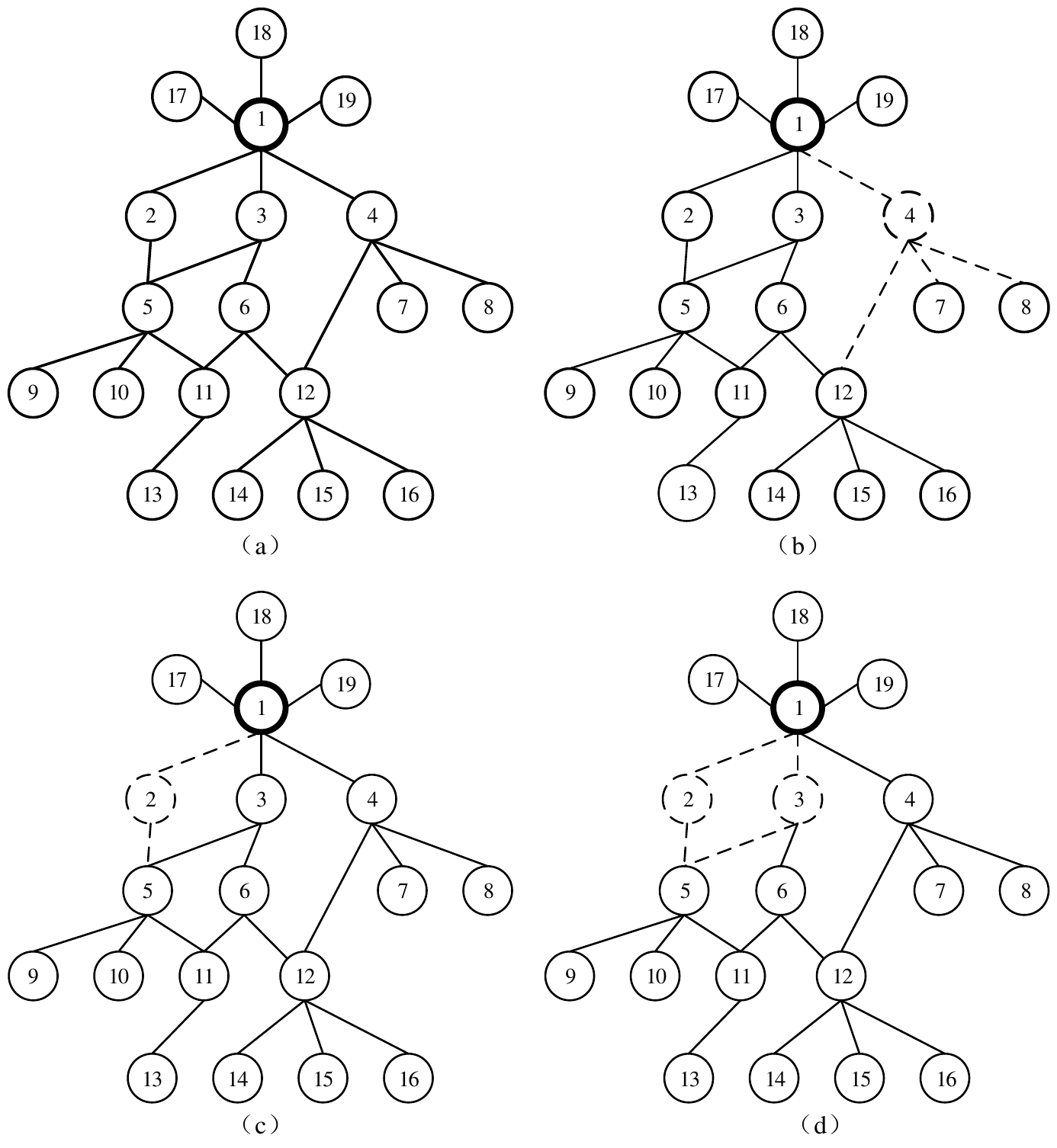}
\caption{Diagram of Internet topology detection based on \textit{traceroute} principle. (a) The original Internet topology, (b) the detected topology after the router $R4$ fails, (c) the detected topology after router $R2$ fails, (d) the detected topology where router $R3$ congests or fails due to the load of router $R3$ being far greater than its capacity, after the router $R2$ failed. The dotted lines and circles denote the removed edges and nodes respectively.}
\label{fig:internet-topo-detection}
\end{figure}

Above all, we analyzed the traceroute principle and the rerouting problem. Next we will discuss the Internet anomaly from the perspective of topology structure detection. In the uncertain network environment, the probed routing path maybe incomplete or inconsistent due to some interference factors. In Fig.\ref{fig:internet-topo-detection}, the changes of network structure are shown by simulating \textit{traceroute} detection process. Fig.\ref{fig:internet-topo-detection}(a) is the original network structure, and an active probe tool is placed at router $R1$. Here we consider two abnormal cases: intentional attack and router failure. When the router $R4$ fails, as shown in the Fig.\ref{fig:internet-topo-detection}(b), not only the paths from $R1$ to $R4$, $R7$ and $R8$ become unreachable, but also the routing paths from $R1$ to $R12$, $R14$, $R15$ and $R16$ will be changed due to rerouting mechanism. In Fig.\ref{fig:internet-topo-detection}(c) the router $R2$ is attacked, all network traffic forwarded by router $R2$ are transferred to router $R3$, such as routing paths $R1-R2-R5$ and $R1-R2-R9$. At this time the router $R3$ maybe into the state of network overload and its network throughout would decrease rapidly. When the router load is greater than its capacity, the router $R3$ doesn't work. The current network topology is described as Fig.\ref{fig:internet-topo-detection}(d). As we can see, the topology changes from (c) to (d) are as a results of Internet cascading failure. We believe that the above problems are more general or even more seriously when the Internet encounters the anomaly events in fact. For example, the Dyn DDoS attack on October 21st 2016 brings some trouble accessing your usual sites and services that happens on the American East Coast primarily but later on the opposite end of the country as well.

\section{Internet Anomaly Detection Method}
\label{sec:internet-anomaly-detection-method}

In this paper, we attempt to detect anomaly behaviors of Internet which has result in the crash of local network, and further made people not accessing the websites or Internet services normally. In general, the Internet behaviors or activities, such as the network device failure, the cable interruption, the abnormal update of router table, and the network traffic breaking out, can be reflected to the real-time Internet topology. However there are billions IP address and tens of billions network devices in the Internet, not all changes are possible to present the phase change of Internet topology. In other word, compared with the giant Internet topology, the small changes cannot influence the Internet work. So we hope to detect the anomaly which have or will result in the local network collapsing and influence human life or work. For the latest American DDoS attack event, the users from different regions unable to access web services due to the Dyn servers' breakdown. We believe that it would be more meaningful if we detect this anomaly events.

Here we hypothesis that all routers adopt the shortest path strategy. In the Fig.\ref{fig:internet-topo-detection}(b), the routing path from $R1$ to $16$ should be $\{R1, R4, R12, R16\}$, if the router $R4$ works normally. However, when the router $R4$ fails, its pre-hop router $R1$ needs to select a new interface to transmit the network packets. According to the shortest path strategy, the new routing path $\{R1, R3, R6, R12, R16\}$ is constructed as the second best. Apparently the new path length is equal to or greater than the old. During the Internet anomaly, the new topology probed will lose some nodes and connections(edges). But the number of missing connections mostly is greater than that of the missing nodes. So the lower the number of network edges is, the lower the network density is, and the larger the network diameter is. In this paper, our purpose is to detect the global Internet anomalies as soon as quickly. So inspired by the study of Xiao\cite{Xiao2008},  which found that the network diameter first increases significantly and then decreases quickly to a relatively small diameter due to the network collapsing thoroughly at bifurcation point in the process of removing more and more network nodes, a novel method of Internet anomaly detection is put forward that focuses on the changes of communication path, such as the mean shortest path length, network effective path and network diameter. Before starting with our method, some definitions about complex network are given.

\begin{definition}
 Given a network $g=(V, E)$, the shortest path length $sp(v,u)$ between node $v$ and $u$ is either the minimum of number of hops in no-weighted network, or the minimum of sum of node- or edge-weights of the path in weighted network. As far as we know, the Internet topology is no-weighted network.
\end{definition}

Where the $V$ and $E$ denote the node set and edge set in network $g$ separately. Hence the $sp(v,u)$ should represent the best information transmission path between node $v$ and $u$, if other conditions are not considered. Then the communication ability of node $v$ can be evaluated by the node diameter $d(v)$(equation\eqref{eq:node-d}) and the node mean shortest path length $p(v)$(equation\eqref{eq:node-spl}). Where the $N(v)$ denotes the set of neighbors of the node $v$. Undoubtedly the $d(v)$ indicates the worst case, and the $p(v)$ is the mean case in process of the node $v$ communicating with others.

\begin{align}
d(v) & =  max\{sp(v,u)|u \in N(v)\} \label{eq:node-d}\\
p(v) & = \frac{1}{|N(v)|}\sum_{u \in N(v)}{sp(v,u)} \label{eq:node-spl}
\end{align}

When the Internet is attacked, the local network crashes and the effective routing paths are changed. As a result, some nodes' diameter and mean shortest path length become bigger. As we all know, the $d(v)$ is greater than $p(v)$, except that all $sp(v,u)$ values are same. Then $d'(v) - d(v) > p'(v) - p(v)$ , where the $d'(v)$ and $p'(v)$ are new values after local network crashes. However the changes of both node metrics could also impact the global network. So we introduce three network metrics to quantify the global network changes.

\begin{definition}
 Given a network $g=(V, E)$, the network diameter $D_{max}$ is defined as
\begin{equation}
D_{max} = \frac{1}{|V|}\sum_{v \in V}{d(v)}
\label{eq:net-md}
\end{equation}
\end{definition}

\begin{definition}
 The effective diameter $D_{effective}$ of the network $g$ is the value of shortest path length within which 90\% of the node pairs are.
\begin{equation}
\begin{split}
& D_{effective} = sp^{\ulcorner 0.9*n\urcorner} ,\\
& sp^{1}\! \leq sp^{2}\! \leq \dots\! \leq sp^{\ulcorner 0.9*n\urcorner}\!\dots\! \leq sp^{n}
\label{eq:net-ed}
\end{split}
\end{equation}
Here let the $n = \frac{|V|*(|V|-1)}{2}$ denote the number of node pairs, and the ${\ulcorner 0.9*n\urcorner}$ represent the minimum integer greater than or equal to $0.9*n$.
\end{definition}

\begin{definition}
 For the network $g$, the average value of all nodes' mean shortest path length is called as the network mean shortest path length $SP$.
\begin{equation}
SP =  \frac{1}{|V|}\sum_{v \in V}{p(v)}\end{equation}
\label{eq:net-sp}
\end{definition}

Based on above definitions, there is the following mathematics relationship: $D_{max} \geq D_{effective} \geq {SP}$. Apparently the inequality $D_{max}-D_{effective} \leq D_{max}-SP$ is also true. We define the network path change coefficient(NPCC):
\begin{equation}
r = \frac{D_{max}-D_{effective}}{D_{max}-SP}
\label{eq:net-r}
\end{equation}
Where the value of $r$ is in the range of 0 to 1.

\begin{figure}[tb]
\includegraphics[width=\linewidth]{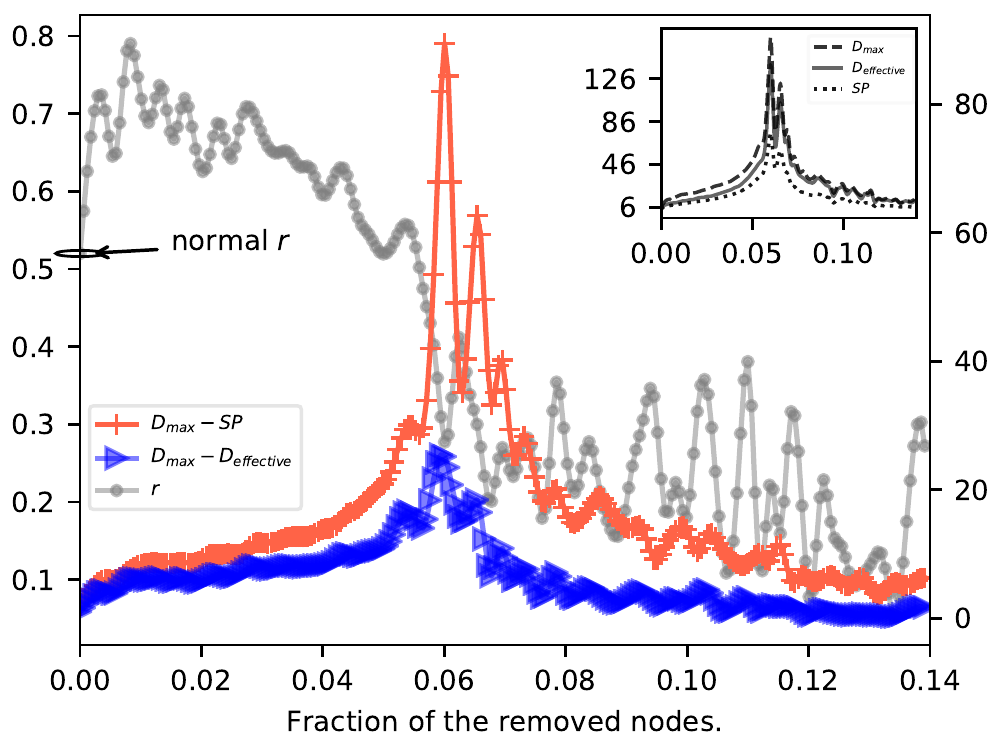}
\caption{The distribution of $D_{max}$, $D_{effective}$, $SP$ and $r$ vs. the fraction of the removed nodes is shown by simulating the intentional attack on BA network with 10000 nodes. The orange lien with $+$ and the green line with $\triangleright$ separately denote the distribution of $D_{max}-SP$ and $D_{max}-D_{effective}$ with with right y-axis, and the gray point is the distribution of $r$ with the left y-axis. The point marked by cycle represents the normal state of the BA network. }
\label{fig:ba-r}
\end{figure}

In this paper, the anomaly events what we study are that they destroy the important infrastructures of Internet, a large number of nodes(e.g. router, server and computer.) and connections disappear from the Internet. As a result, people in local area can not access Internet. Intuitively, we would think that the value of $r$ increases and is close to 1 gradually. That is because $d'(v) - d(v) > p'(v) - p(v)$, when the network is attacked or fails, then $D'_{max}-D_{max} > D_{effective} - D_{effective} > SP'-SP$ is true. In other word, the change rate of $D_{max}$ is biggest, and that of $SP$ is smallest. But the study results in Ref.\cite{Xiao2008} show that the network structure will experience one process from quantity changing to quality changing during Internet anomaly occurring. It is that the value of the $D_{max}$ increases first, and then decreases quickly after the enough nodes are removed. So the $r$ should decreases at someone state of Internet yet. In order to verify our analysis, a simulation experiment of intentional attack on BA network is implemented. From the Fig.\ref{fig:ba-r}, it can be seen that the value of $r$ increases first and then decreases indeed. Start with a Internet attack, the $D_{effective}$ and $SP$ of the network grow slowly compared with the $D_{max}$, because the number of removed nodes is too few that doesn't impact the global Internet structure. But this behavior will change the worst case, i.e. $D_{max}$. So the $r$ is growth. As the number of removed nodes increases, more and more the shortest path length of node pairs $sp(v,u)$ are increased. The result is the $D_{effective}$ growth correspondingly. Furthermore the $D_{max}-SP$ is more quick than the $D_{max}-D_{effective}$. As the Fig.\ref{fig:ba-r} shown, the value of the $r$ starts to decline after the 5.39\% of nodes are deleted. At the point of 6.03\% removed nodes, the $D_{max}$, $D_{effective}$ and $SP$ arrive the maximum value of this network. The value of the $r = 0.3489$ has already lower than the initial value $0.5201$. At this time, we also can say that the Internet structure undergoes a qualitative changing. The local network collapsing lets this area form a big black hole in Internet. Even the surround nodes could be influenced due to the cascading failure.

% \begin{equation}
% \begin{split}
% r & = \frac{D_{max} - D_{effective}}{D_{max} - SP} \\
% & = \frac{ \frac{1}{|V|}\sum_{v \in V}{d(v)} - d^{\ulcorner 0.9*n\urcorner +1}}{\frac{1}{|V|}\sum_{v \in V}{d(v)} - \frac{1}{|V|}\sum_{v \in V}{\alpha(v)} } \\
% % & = \frac{ \frac{1}{|V|}\sum_{v \in V}{max\{sp(v,u)|u \in N(v)\}} - d^{\ulcorner 0.9*n\urcorner +1}}{\frac{1}{|V|}\sum_{v \in V}{max\{sp(v,u)|u \in N(v)\}} - \frac{1}{|V|}\sum_{v \in V}{\frac{1}{|N(v)|}\sum_{u \in N(v)}{sp(v,u)}} } \\
% & = \frac{ \frac{1}{|V|}\sum_{v \in V}{max\{sp(v,u)|u \in N(v)\}} - d^{\ulcorner 0.9*n\urcorner +1}}{\frac{1}{|V|}\sum_{v \in V}{max\{sp(v,u)\} - \frac{1}{|N(v)|}\sum{sp(v,u)}}}
% \end{split}
% \end{equation}

In a word, the value of the $r$ will increase first and then decrease, as shown in Fig.\ref{fig:ba-r}, when the network is attacked. So for a definite complex network with the initial NPCC $r_0$, if current $r$ is greater than $r_0$, the Internet anomaly would have happened but its structure has not been changed qualitatively; then the $r$ starts to decrease when the attacked network is close to or at the point of phase transformation due to quality changing; and finally the $r$ is far below the $r_0$. However, the complex networks are a dynamical system by adding or removing the nodes and edges. In the study of Ref.\cite{Albert2002}, the authors found that the network diameter vs. the number of network nodes follows the logarithmic distribution, i.e. $D_{max} \sim lg|V|$. Thus the normal fluctuations of network structure is very small, compared with the changes caused by the abnormal events. In this paper, an redundancy range is defined,
\begin{equation}
\eta = [r_0 - \tau, r_0 + \tau]
\end{equation}
so as to present the network fluctuation behavior. Where the $\tau$ denotes the fluctuation parameter of the normal value. Furthermore it can say that the network is abnormal, if the $r \notin \eta$, or vice versa. Under the ideal conditions, the $r$ of a network should be a fixed value. But for a dynamic network, it should be in a acceptable fluctuation range. In general, the current network state is related with the previous. Just like the Fibonacci sequence, every number is the sum of the two preceding ones, i.e. $f(i) = f(i-1)+f(i-2)$. Inspired by it, the normal domain at time tick $i$ is constructed as following:
\begin{equation}
f(i) = \eta(i-1) + \eta(i-2) + \dots + \eta(i-k)
\label{eq:f-func}
\end{equation}
Here the $\eta$ follows the below relationship: $\eta(i) + \eta(j) = \eta(i) \cup \eta(j), \forall{i,j} \in \{ 1, 2, \dots, n\} $. Furthermore the normal domain of Internet is $F=\{f(1), f(2), \dots, f(n)\}$

The notation $k$ implies that the current network state is related with previous $k$ network states. Every network state $r$ is defined by an redundancy range $\eta$ to decrease the impact of the normal fluctuation of Internet. So the anomaly detection results are decided by the $k$ and $\tau$. According to the hoeffding's inequality\cite{Hoeffding1963},  the $\tau$ is given by the following equation:
\begin{equation}
\tau = \mu + \lambda \sigma
\label{eq:f-fun-2}
\end{equation}
Where $\mu$ denotes the mean and $\sigma$ is the standard variation of the previous $r$. And the notation $\lambda$ is the quantile of the normal distribution corresponding to the given confidence interval $\theta$\cite{meeker2014statistical}.

\begin{figure}[b]
\begin{center}
\includegraphics[width=\columnwidth]{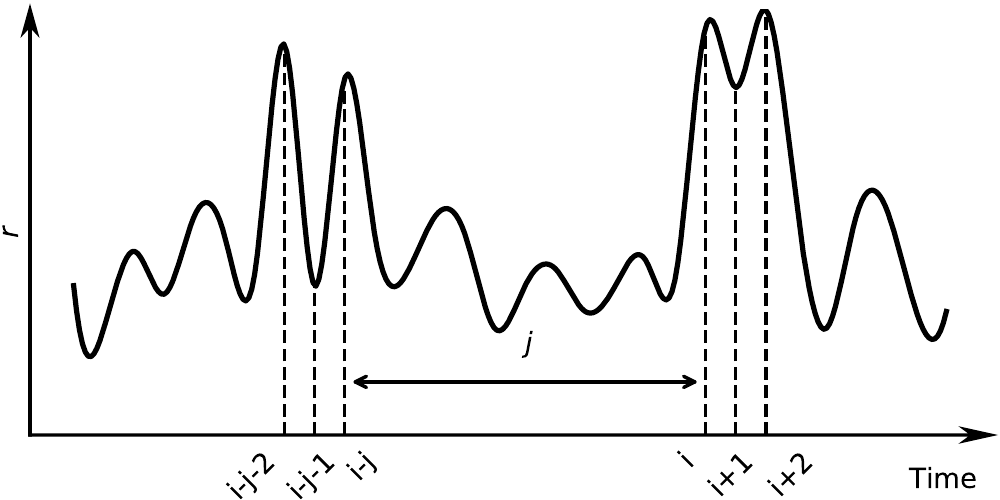}
\end{center}
\caption{Diagram of the $r$ evolution.}
\label{fig:net-diameter}
\end{figure}

Next, we will discuss the relationship between $k$ and $f$ which is described by the Fig.\ref{fig:net-diameter}. When $k = 1, f(i)=\eta(i-1)$, the equation\eqref{eq:f-func} tells us that the current result is related to the preceding network only. So the decision function $f$ can find the abnormal points $i-j-2$, $i-j-1$, $i-j$, and $i$. That is because the values of $r(i-j-2)$, $r(i-j-1)$, $r(i-j)$ and $r(i)$ will exceed the normal range of decision function $f(i-j-2)=\eta(i-j-3)$, $f(i-j-1) = \eta(i-j-2)$, $f(i-j)=\eta(i-j-1)$ and $f(i)=\eta(i-1)$ separately. In fact, the network $i-j-1$ is normal, and the $i+1$ and $i+2$ are abnormal. Although the $k=2, f(i) = \eta(i-1) + \eta(i - 2)$, can avoid the error $i-j-1$, it will regard the network $i-j$ as normal state. In order to avoid this errors, a restrictive condition is added to the equation\eqref{eq:f-func}, that is the previous $k$ network state are normal state. About the problem of the value of $k$, we have a study in more detail at the Section\ref{sec:results}

\section{Experimental Results}
\label{sec:experimental-results}

\subsection{Datasets}
\label{sec:datasets}
In this paper, the Internet topology datasets are used to verify the effectiveness and performance of our model. Since the Internet topology is a dynamical complex network who faces the external intrusion and internal failure all the time. Fortunately there have been some Internet detection projects to capture the AS-level, router-level and IP-level Internet topology structure such as Routeviews, RIPE NCC, Caida Ark, and DIMES.

In order to study abnormal behaviors of the Internet, the Internet topology dataset from CAIDA's Ark project\cite{caida2017} was used which includes three time periods of topology data in 2011's Japanese earthquake, 2014's BGP table hitting 512k and 2016's American DDoS attack. Apparently those anomaly events are involved with not only network device failure, but also Internet intentional attacks. Noted that all time in this paper are the zero-timezone time (UTC).

First, Internet active probe data of each of anomaly event was downloaded for three days, which covered the time before the event will occur, when it is occurring, and after it occurred. Second, these data records were filtered in fixed time window, i.e. starting from two days before and end with two days after the event happens. During data processing, we found that the length of a few routing paths are greater than 128, however, Ref.\cite{Golkar2014} shows that 65\% Internet paths length are in range of 7 to 12 and a few paths are longer than 18 hops. This is because that the Scamper\cite{Luckie2010}, as the Ark probe tool, does note control the Time-To-Live flag of IP packet effectively. So we had removed the records that its path length is greater than 64 in this paper. Next, we choose the sample rate of 10 minutes for this dataset (with 720 time ticks) as lower sample rates result in excess number of change points due to large fluctuations in network structure over small time periods, on the other hand, higher sample rates obscure the true positive(i.e., actual changes). Finally, The Internet topologies were extracted from the routing paths of sampling files.

\begin{figure*}[t!]
\begin{center}
\includegraphics[width=0.8\linewidth]{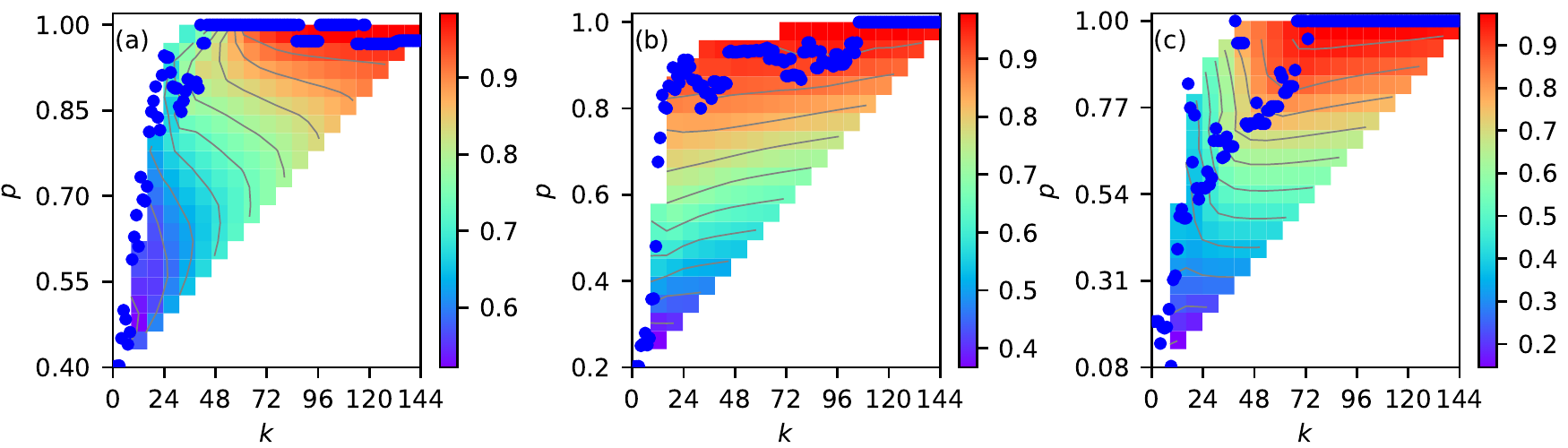}
\caption{The relationship among the $k$(Eq.\eqref{eq:f-func}), precision $p$ and $F1$. The colorbar denotes the range of $F1$ with different $k$ and $p$. However the blue points describe the relationship between $k$ and $p$. Here three sub-figures are that of Japan earthquake, BGP hitting 512K and American DDoS attack respectively. }
\label{fig:k-p-f}
\end{center}
\end{figure*}

\subsection{Results}
\label{sec:results}

In order to quantitatively evaluate the detection performance, the widely used criteria including the detection accuracy $\alpha$ , detection precision $p$ and detection recall $r$ of anomaly events are introduced in this paper. First four notations are defined: $TP$ denotes the number of abnormal points detected which are in the anomaly event yet, $FP$ is the number of abnormal points while they are normal in fact, $TN$ represents the number of detected normal points and they are normal in data yet, $FN$ is the number of normal points while they are abnormal in fact. The metrics are defined as follows:
\begin{equation}
\begin{split}
\alpha & = \frac{TP+TN}{TP+FP+TN+FN} \\
p & = \frac{TP}{TP+FP} \\
r & = \frac{TP}{TP+FN} \\
F_b &= \frac{( 1 + b^2)*p*r}{{b^2}*p+r} \\
\end{split}
\end{equation}
Where the detection accuracy $\alpha$ is the ratio of correct results detected among sampled networks, detection precision $p$ is the correct ratio among the abnormal points which are determined by our model, and detection recall $r$ is the ratio of correct abnormal points over total true abnormal points in testing data. Under the ideal conditions, it would be better if the values of $p$ and $r$ are bigger. In fact, the $p$ is higher with lower $r$, vice versa. For considering the precision $p$ and the recall $r$ of the test, the $F_b$ score can be interpreted as a weighted average of the precision and recall, where a $F_b$ score reaches its best value at 1 and worst at 0.

First, we will discuss a problem what the value of $k$ is best in our method. In the equation\eqref{eq:f-func}, the $k$ suggests that the current network state should be related with multiple preceding network state. In fact, a study of the self-similarity of network traffic\cite{Crovella1997} has found that wide-area traffic levels follow 24-hour patterns. But the Internet topology reflects the connectivity among network devices, i.e. the routing paths. Although the normal network traffic can not influence the Internet topology structure, the abnormal network traffic will trigger the network load balancing strategy and even cause the cascading failure. So the normal network state should be consistent with that of the previous, or be in acceptable range of fluctuation. In our experimental, the $k$ shows the correlation with the Internet over first $k*10$ minutes. In order to study the value of $k$, we have analyzed the distribution among the $k$, precision $p$ and $F1$ as shown in Fig.\ref{fig:k-p-f}. The results show that in the beginning the values of $p$ and $F1$ grow with the increasing $k$. Then they are close to 1 after about $k=60$ which be depicted by the red area and the top contour line in figure. Noted that, under the same of anomaly detection precision, e.g. $p>0.8$ \& $F1>0.8$, the value of $k$ in BGP hitting 512K event, about 16, is far less than that in other events. It denotes that the changes of Internet topology are larger because this event occurs in the Internet backbone. This changes was verified in Fig.\ref{fig:results} yet. Meanwhile the $k$ of Japan earthquake is less than that of American DDoS attack. This result indicates that the more serious the anomaly event and the wider its impact, the smaller the value of $k$.

Furthermore we analysis the performance of our method when the $k$ is 12, 24, 36, 48 and 60. According to the results of Table\ref{tab:a-p-r-f}, it can be seen that the detection accuracy of three anomaly events is not less than 90\% , but the values of the precision $p$ are less than 60\% in some results. For instance the American DDoS attack with $k=12$, that is because more than 90\% of sampling points belong to the $TN$ points. It also indicates that the Internet topology structure anomaly is intermittent and not continuous, even though the Internet could be damaged by the strong earthquake disaster. This also implies that the Internet is a robust system. When the Internet is attacked or fails, on one hand the router will choose new forwarding path by the routing balancing strategy after one path fails, and on other hand the Scamper always sends probe packets by the keeping IP address list.

According to observing the values of $F1$, we found that the performance of anomaly detection of our method arrives a high level when $k = 60$, that is 0.9118, 0.9545 and 0.9268 respectively. However the value of $F1$ in American DDoS attack is less than the other two, when $k<=48$. The main reason is that the attack occurs in American and doesn't have a impact to the global Internet. But the Japan earthquake had influenced the communication in the Asia-Pacific and the massive route leak had caused significant network problems for the global routing system. It infer that more serious the anomaly event, the smaller the $k$.

\begin{table}[b]
\begin{center}
\caption{\label{tab:a-p-r-f} The performance of Internet anomaly detection with $k =12, 24, 36, 48 $ and $ 60$ based on our method.}
\begin{tabular}{c c c c c c }
\hline
\hline
$k$ & Events & $\alpha$ & $p$ & $r$ & $F_1$ \\
\hline
 \multirow{3}{*}{12} & Japan earthquake & 0.9056 & 0.6119 & 0.4940 & 0.5467 \\
& BGP table hitting 512k & 0.9681 & 0.6761 & 1.0000 & 0.8067 \\
& American DDoS attack & 0.9333 & 0.3939 & 0.3171 & 0.3514 \\
\hline
 \multirow{3}{*}{24} & Japan earthquake & 0.9653 & 0.9459 & 0.6034 & 0.7368 \\
& BGP table hitting 512k & 0.9903 & 0.9123 & 0.9630 & 0.9369 \\
& American DDoS attack & 0.9625 & 0.5556 & 0.3448 & 0.4255 \\
\hline
 \multirow{3}{*}{36} & Japan earthquake & 0.9792 & 0.9024 & 0.7708 & 0.8315 \\
& BGP table hitting 512k & 0.9847 & 0.8333 & 0.9804 & 0.9009 \\
& American DDoS attack & 0.9778 & 0.6667 & 0.6667 & 0.6667 \\
\hline
 \multirow{3}{*}{48} & Japan earthquake & 0.9792 & 1.0000 & 0.6341 & 0.7761 \\
& BGP table hitting 512k & 0.9958 & 0.9302 & 1.0000 & 0.9639 \\
& American DDoS attack & 0.9861 & 0.7273 & 0.8000 & 0.7619 \\
\hline
 \multirow{3}{*}{60} & Japan earthquake & 0.9917 & 1.0000 & 0.8378 & 0.9118 \\
& BGP table hitting 512k & 0.9944 & 0.9333 & 0.9767 & 0.9545 \\
& American DDoS attack & 0.9958 & 0.8636 & 1.0000 & 0.9268 \\
\hline
\hline
\end{tabular}
\end{center}
\end{table}

\begin{figure}[b]
\begin{center}
\includegraphics[width=0.8\linewidth]{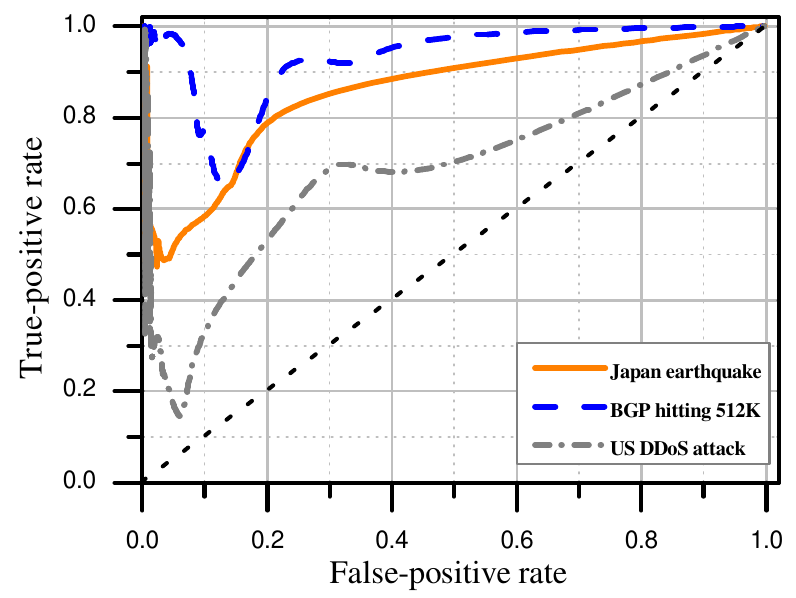}
\caption{The ROC curve of Internet anomaly detection.}
\label{fig:roc-res}
\end{center}
\end{figure}

To eliminate the influence for the detection performance from the $k$, the receiver operating characteristic(ROC) criteria is introduced for the further comparison among three events, which is shown in Fig.\ref{fig:roc-res}. It shows that the ROC curve of BGP hitting 521k is above the one of Japan earthquake, and the values of both are always greater than that of American DDoS attack. Furthermore it implies that the performance of Internet anomaly detection is proportional to the scope and severity of the event.

\section{Discussion}

\begin{figure}[b]
\begin{center}
\includegraphics[width=\linewidth]{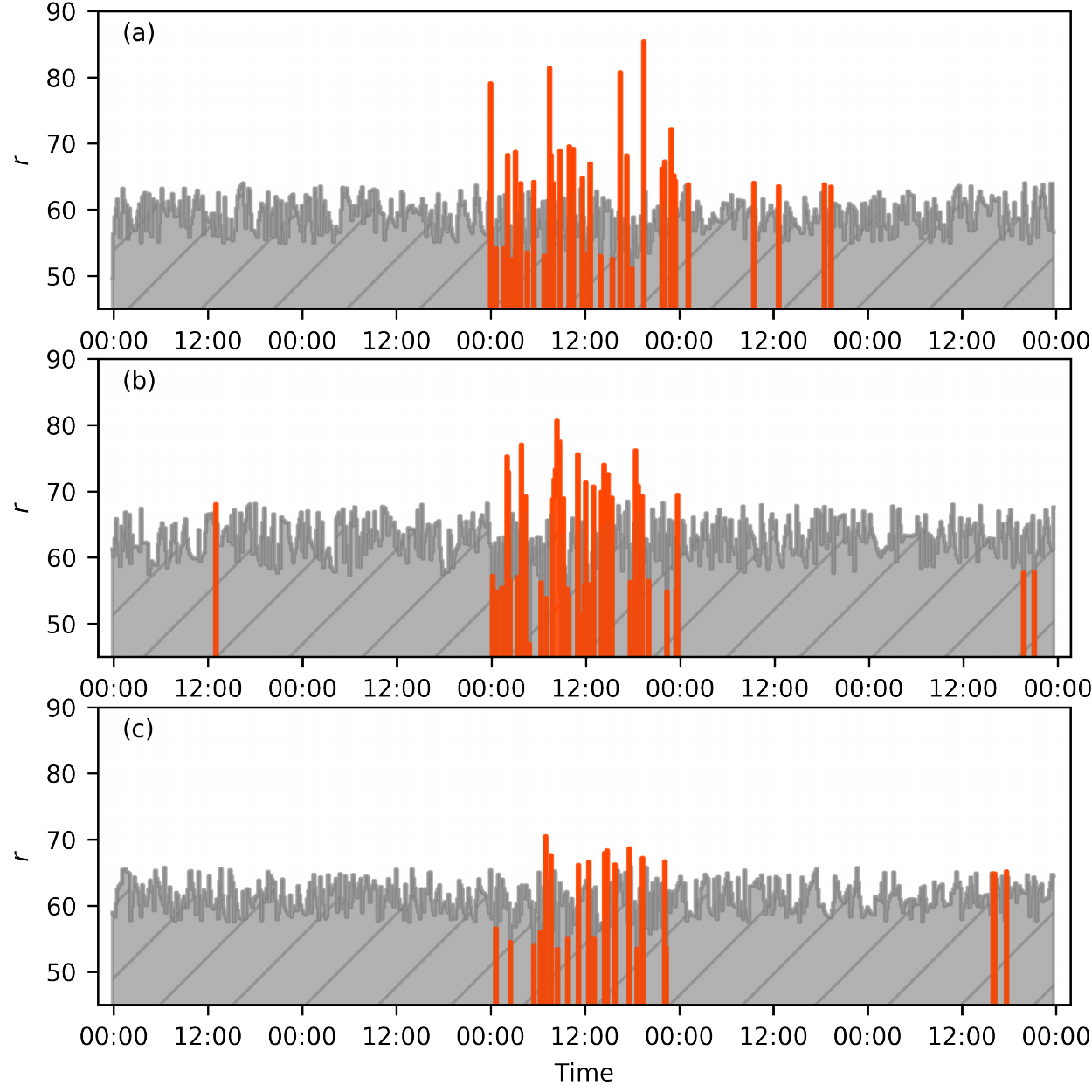}
\caption{The anomaly detection results of our model. (a) Japan earthquake event, (b) BGP hitting 512K event, and (c) US DDoS attack event. Where the bars denote the distribution of network diameter vs. time, and the grey bar represents the normal sampled network and the red bar is the abnormal. }
\label{fig:results}
\end{center}
\end{figure}

In this subsection, the detection results of every Internet anomaly event, as the figure\ref{fig:results} shown, will be discussed in more detail.

First for the Japan earthquake event, we think that it should belong to the network failure, i.e. a large number of the routers, computers, servers and other network devices, especially the backbone routers which are as exchange points to forward data packets, have failed due to the power or cables interruption. In order to evaluate our work accurately, we have collected some news reports and papers \cite{Cho2011,Liu2012}. These documents will better help us to label the abnormal points of testing data. As far we know, the main earthquake of magnitude 9.0 was preceded by a number of large foreshocks starting two days earlier, and hundreds of aftershocks continuing for months. Just after the main earthquake at 5:46 on March 11th 2013, a series of aftershocks followed, including ones with more than M7, an M7.4 at 6:09, an M7.7 at 6:16, and an M7.5 at 6:26. So the Fig.\ref{fig:results}(a) shows that we detect the continuous Internet abnormal behaviors on the morning of March 11th 2013. The quake and tsunami brought about multiple submarine cables cut, seriously impacted the intra-Asia and trans-Pacific communications and Internet access. On the evening of March 13th, most of the backbone connectivities were restored\cite{Cho2011}. There is no abnormal detected in the Fig.\ref{fig:results}(a). And business service activities as well as customer service and support work started two days later. So the Japanese network experienced limited damage because rerouting traffic by intact cables lines.

Next Internet anomaly event--BGP hitting 512K in old router--will be analyzed. What happened was the Internet's growth exceeding the default configuration limit of certain models of network switching equipments. After this event happened, Geoff Huston had made a detailed report from the perspective of routing updates including the activities of announced prefixes and withdraw prefixes, and the variation of BGP FIB size\cite{bgp2017,huston2014s}. Based on them, we label the abnormal points of testing data and to discuss our detection results more in depth. The initial explosion of this event was in 7:40 $\sim$ 8:10 on August 12th. However the company LiquidWeb reported on Twitter at 5:19 that the problem first appeared to be the result of a "large network provider is performing maintenance". In fact, around 2:00 on August 12th, the BGP FIB size is very closely the network device capacity. But from the growth trend of BGP FIB size, the Internet routing table had passed 512K stably at 12:00 on August 12th, except that it happens a larger fluctuation around 18:00 on August 12th. So it can bee seen in Fig.\ref{fig:results}(b) that this anomaly event was mainly detected in four time periods, just like what we analyze.

The third anomaly event is a large DDoS attack against Dyn, one of the leading authoritative DNS providers. The attack started around 11:00 on October 21th 2016 and lasted for hours, severely hurting the reachability of big name sites like Twitter, GitHub and PayPal. According to the technique reports from Dyn, we give the timeline of the massive DDoS attacks, as shown in Table \ref{tab:ddos-timeline}. From the Dyn's report, the attack had been resolved at 22:17.  As far we know, this attack is a botnet coordinated through a large number of Internet of Things-enabled devices, observing 10 of millions of discrete IP addresses which Dyn wrote, that had been infected with Mirai malware. So it will take some time stopping the attack behaviors of all devices, and even more time making Internet system recover to the stable state from the attack.
\begin{table}[tb]
\begin{center}
\caption{\label{tab:ddos-timeline} Timeline of American DDoS attack on October 21th 2016.}
\begin{tabular}{ p{0.12\columnwidth} |p{0.83\columnwidth}}
\hline
\hline
Time & Remarks \\
\hline
11:10 &
Dyn began experiencing a DDoS attack. Some users may experience the increased DNS query latency and delayed zone propagation during this time.
\\
12:45 &
The attack was impacting the East Coast of US, and Internet users directed to Dyn server were unable to reach some of websites.
\\
14:00 &
Dyn mitigated the attack and restored the service to customers.
\\
15:50 &
The Dyn had begun monitoring and mitigating the second DDoS attack.
\\
18:00 &
This second wave had impacted the Internet in the US and some customer would have seen extended latency delays during this time. The Internet service was substantially restored at approximately now.
\\
18:23 &
Dyn Managed DNS advanced service monitoring was currently experiencing issues and resolved half an hour later.
\\
20:30 &
There was residual impact from additional sources.
\\
22:17 &
This event had been resolved.
\\
\hline
\hline
\end{tabular}
\end{center}
\end{table}

\section{Conclusions and Future Work}
\label{sec:conclusions}

In this paper, we have proposed a method of detecting Internet anomaly behaviors by analyzing the dynamics of real-time Internet topology. According to studying the network path change ratio between network diameter, network effective path and network mean shortest path under abnormal conditions, such as network devices failure or intentional attack,  The anomaly detection method from the perspective of monitoring the dynamics of global IP-level Internet topology structure, instead of network traffic. As we all know, the routing paths is incomplete or rerouted because of network failure or Internet cascading failure effect. Thus a new complex network metric NPCC $r$ is proposed to quantify the network path change ratio under normal and abnormal behaviors. We found that the NPCC $r$ first increases when the anomaly start to happen, then decreases continuously to be much lower than the initial $r$ until the network structure reaches the phase transaction point--from quantity changing to quality changing. However for the Internet as a complex dynamic system, the NPCC $r$ is not a fixed value, thus we has constructed the decision function of $r$ inspired by the Fibonacci Sequence that the current Internet NPCC is related to previous $k$ values, and further a normal domain $f$ is used to determine the current Internet to be abnormal, if $r > f$. Finally, we use our method to three Internet anomaly events: 2011's Japanese earthquake, 2014's BGP table hitting 512k and 2016's American DDoS attack. The experimental results show that the detection accuracy of our method are 97.92\%, 98.47\% and 97.78\% respectively, when the $k=36$, i.e. the current $r$ is related with the that about the last 6 hours. We can say that our method can detect Internet anomaly event effectively. For the detection precision with different $k$ as shown in Fig.\ref{fig:k-p-f}, we found that the bigger the $k$ value, the better the detection precision. In Fig.\ref{fig:roc-res}, it suggests that our method has a better detection accuracy in the wider Internet abnormal event. Compared with the methods based on network traffic, our method has following advantages: (1) monitor the global Internet operations state, (2) capture the local collapsed subnetwork because of Internet anomaly, and (3) don't need the special device to mirror the network traffic.

In the future works, we will develop a method to locate the location of Internet anomaly , after it is detected. So it would be very meaningful that predict the evolving trend of Internet anomaly and assess the extent of hazard, if we locate the faulty LAN network.

\section*{Acknowledgement.} This work is based upon work supported by the
Liaoning Science and Technology Project under grant number 2015401039.

\bibliographystyle{ieicetr}% bib style
\bibliography{refs}% your bib database
%\profile{}{}
%\profile*{}{}% without picture of author's face
% \profile{Hanako Denshi}{received the B.S. and
% M.S. degrees in Electrical Engineering from
% Denshi Institute of Technology in 1997
% and 1999, respectively. ...}
\end{document}